\documentclass[showpacs,10pt,twocolumn,prl]{revtex4-1}

\usepackage{amsmath}
\usepackage{amssymb}
\usepackage{graphicx}
\usepackage{amssymb}
\usepackage{graphics}
\usepackage{epsfig}
\usepackage{CJK}
\usepackage{color}
\usepackage{soul}

\begin{document}

\begin{CJK*}{GBK}{Song}
\title{Physical properties of the layered $f$-electron van der Waals magnet Ce$_2$Te$_5$}
\author{Yu Liu, M. M. Bordelon, A. Weiland, P. F. S. Rosa, S. M. Thomas, J. D. Thompson, F. Ronning, and E. D. Bauer}
\affiliation{MPA-Q, Los Alamos National Laboratory, Los Alamos, New Mexico 87545, USA}
\date{\today}

\begin{abstract}
We report a detailed study of the magnetic, transport, and thermodynamic properties of Ce$_2$Te$_5$ single crystals, a layered $f$-electron van der Waals magnet. Four consecutive transitions at $\sim$ 5.2, 2.1, 0.9, and 0.4 K were observed in the $ac$-plane electrical resistivity $\rho$(T), which were further confirmed in specific heat $C_\textrm{p}$(T) measurements. Analysis of the magnetic susceptibility $\chi$(T), the magnetic-field variation of $\rho$(T), and the increase of the first transition temperature ($T_\textrm{c} \sim$ 5.2 K) with applied magnetic field indicates ferromagnetic order, while the decrease of the other transitions with field suggests different states with dominant antiferromagnetic interactions below $T_2 \sim$ 2.1 K, $T_3 \sim$ 0.9 K, and $T_4$ = 0.4 K. Critical behavior analysis around $T_\textrm{c}$ that gives critical exponents $\beta = 0.31(2)$, $\gamma = 0.99(2)$, $\delta = 4.46(1)$, $T_\textrm{c} = 5.32(1)$ K indicates that Ce$_2$Te$_5$ shows a three-dimensional magnetic critical behavior. Moreover, the Hall resistivity $\rho_{\textrm{xy}}$ indicates that Ce$_2$Te$_5$ is a multi-band system with a relatively high electron mobility $\sim 2900$ cm$^2$ V$^{-1}$ s$^{-1}$ near $T_\textrm{c}$, providing further opportunities for future device applications.
\end{abstract}
\maketitle
\end{CJK*}

\section{INTRODUCTION}

Layered van der Waals (vdW) materials have attracted widespread attention due to the exotic quantum states they exhibit, such as correlated insulating, ferromagnetic, and superconducting states in ``twisted'' bilayer graphene \cite{CaoY1,CaoY2,Matthew}, or quantum criticality in twisted transition metal dichalcogenides \cite{Ghiotto}. The discovery of intrinsic long-range magnetic order in monolayer CrI$_3$ and bilayer CrGeTe$_3$ has opened up new avenues of research into magnetism in the two-dimensional (2D) limit \cite{Huang,Gong} as well as integration of 2D magnetic layers for control of magnetism by gating or other electrical means in devices \cite{Ralph2019,Burch2018,Song2018,ACSNANO,YL,YL001,YL002}. Accordingly, several other $3d$-electron vdW magnets, such as FePS$_3$, Fe$_3$GeTe$_2$, VSe$_2$, VI$_3$, CrTe$_2$, and MnSe$_2$, have been extensively investigated \cite{Lee,Deng,Bonilla,Zhang,Hara,VI3,Fe3GeTe2}. In contrast, very few $f$-electron vdW magnets have been studied (e.g., CeSiI \cite{CeSiI}, EuC$_6$ \cite{EuC6_1,EuC6_2,EuC6_3}, and GdTe$_3$ \cite{Lei}). Many $f$-electron materials exhibit significant hybridization between the $f$-electrons and conduction electrons, leading to highly correlated quantum states with narrow $f$-bands near the Fermi level; thus, 2D $f$-electron vdW materials may be highly tunable with modest amounts of pressure, uniaxial strain, or magnetic field, making them promising candidates for discovering and exploring unusual quantum states.

The family of rare-earth telluride RTe$_x$ (R = rare-earth element; $x$ = 2, 2.5, and 3) adopts a layered crystal structure, consisting of square planar Te layers and corrugated RTe slabs. The RTe slabs are semiconducting and responsible for magnetism, while the Te layers form 2D conducting bands; thus RTe$_x$ exhibits highly anisotropic transport and magnetic properties \cite{Min,Kwon,Shin,Iyeiri,Ru,Ru1}. Among this series, CeTe$_2$ and CeTe$_3$ crystallize in the layered Cu$_2$Sb-type tetragonal (space group: $P4/nmm$) structure and NdTe$_3$-type weakly orthorhombic (space group: $Cmcm$) structure, respectively, with localized Ce$^{3+}$ magnetic moments. CeTe$_2$ contains a single layer of Te and undergoes an antiferromagnetic (AFM) transition at $T_\textrm{N}$ = 4.3~K \cite{Kwon1,Min1,Jung,Jung1}. At 2 K, a metamagnetic transition to a field-induced ferromagnetic (FM) state with an easy $c$ axis occurs at a small magnetic field of 0.06 T. The resistivity shows a sharp peak at $T_\rho$ = 6 K well above $T_\textrm{N}$ with a large negative magnetoresistance (MR) \cite{Jung,Jung1,Jung2,Kasuya,Min2} arising from magnetic-polaron and/or short-range FM ordering. Neutron diffraction measurements indicate a down-up-up-down AFM configuration along the $c$ axis with FM Ce double layers above and below the Te layer in CeTe$_2$ \cite{Park,Park1,Stowe,Liu1,Shim}. CeTe$_3$ contains double layers of Te connected via weak vdW force; it exhibits two AFM transitions at $T_{\textrm{N1}}$ = 3.1 K and $T_{\textrm{N2}}$ = 1.3 K with non-parallel easy axes that are perpendicular to the layer stacking direction, i.e., strongly easy-plane character \cite{Deguchi,Zocco,Okuma,Watanabe}.

Layered Ce$_2$Te$_5$ can be considered as a combination of CeTe$_2$ and CeTe$_3$ (Fig. 1), consisting of alternating single and double Te layers stacked along the $b$ axis of the orthorhombic unit cell and separated by CeTe slabs \cite{Chen}. Ce$_2$Te$_5$ crystallizes in a weakly orthorhombic structure, similar to CeTe$_3$, with two Ce sites either adjacent to the double layers of Te (Ce1) or to the monolayer of Te (Ce2). Chen $et$ $al.$ reported three magnetic transitions at 5.1, 2.3, and 0.9 K in Ce$_2$Te$_5$ single crystals \cite{Chen}.

In this study we report the physical properties of single crystals of Ce$_2$Te$_5$, including magnetic susceptibility, magnetization, specific heat, and longitudinal and Hall resistivity measurements. An additional magnetic transition at $\sim$ 0.4 K was observed, where the resistivity features a weak kink and the specific heat exhibits a peak. When magnetic field is applied along the $b$ axis, the resistivity shows that the first transition $T_\textrm{c} \sim 5.2$ K broadens and shifts to higher temperatures, consistent with FM ordering; however, the second and third transitions at $T_2$ = 2.1 K and $T_3$ = 0.9 K move to lower temperatures. The critical exponents obtained around $T_\textrm{c}$ indicates that Ce$_2$Te$_5$ shows a three-dimensional magnetic critical behavior. Furthermore, the Hall effect suggests that Ce$_2$Te$_5$ is a multi-band system with a relatively high electron mobility around $T_\textrm{c}$.

\section{METHODS}

\subsection{Experimental details}

Single crystals of Ce$_2$Te$_5$ with typical dimensions of $3\times3\times0.1$ mm$^3$ were grown by a RbCl/LiCl flux \cite{Chen}. The crystallographic structure of Ce$_2$Te$_5$ was verified at room temperature by a Bruker D8 Venture single-crystal X-ray diffractometer equipped with Mo radiation. X-ray diffraction analysis shows that Ce$_2$Te$_5$ crystallizes in the orthorhombic space group $Cmcm$ (No. 63) with lattice parameters $a \sim c \approx$ 4.42 {\AA} and $b \approx$ 44.17 {\AA}, in agreement with a previous report \cite{Chen}.

The magnetization was measured in a Quantum Design Magnetic Property Measurement System (MPMS) from 2 to 350 K up to magnetic fields $\mu_0H$ = 6 T and $\mu_0$ is magnetic permeability in vacuum. For critical analysis, the reported internal magnetic field ($\mu_0H_\textrm{int}$) has been corrected, $\mu_0H_\textrm{int} = \mu_0H - NM$, where $\mu_0H$ is the applied magnetic field, $M$ is the measured magnetization, and $N \sim 0.95$ is the demagnetization factor. The specific heat was measured using a Quantum Design Physical Property Measurement System (PPMS) from 0.35 to 20 K that utilizes a quasi-adiabatic thermal relaxation technique. The longitudinal and Hall resistivity were measured in a PPMS using standard four-probe configurations with the current flowing in the $ac$-plane and the magnetic field applied along the $b$-axis. The Hall resistivity $\rho_{\textrm{xy}}(\mu_0H)$ was calculated by the difference of transverse resistivity measured at positive and negative fields, i.e., $\rho_{\textrm{xy}}(\mu_0H) = (\rho_{\textrm{H+}}-\rho_{\textrm{H-}})/2$, so as to effectively eliminate the longitudinal resistivity contribution due to voltage probe misalignment.

\subsection{Scaling analysis}

A second-order phase transition around the Curie temperature $T_\textrm{c}$ is characterized by a set of interrelated critical exponents $\beta$, $\gamma$, $\delta$ and a magnetic equation of state \cite{Stanley}. The critical exponents $\beta$ and $\gamma$ are associated with the spontaneous magnetization $M_\textrm{s}$ and the inverse initial susceptibility $\chi_{\textrm{ini}}^{-1}$, below and above $T_\textrm{c}$, respectively, while $\delta$ is the critical isotherm exponent. The definitions of $\beta$, $\gamma$, $\delta$ from magnetization measurement are given below:
\begin{equation}
M_\textrm{s} (T) = M_0(-\varepsilon)^\beta, \varepsilon < 0, T < T_\textrm{c},
\end{equation}
\begin{equation}
\chi_{\textrm{ini}}^{-1} (T) = (\mu_0h_0/m_0)\varepsilon^\gamma, \varepsilon > 0, T > T_\textrm{c},
\end{equation}
\begin{equation}
M = D(\mu_0H_\textrm{int})^{1/\delta}, T = T_\textrm{c},
\end{equation}
where $\varepsilon = (T-T_\textrm{c})/T_\textrm{c}$ is the reduced temperature, and $M_0$, $\mu_0h_0/m_0$ and $D$ are critical amplitudes \cite{Fisher}.

The magnetic equation of state in the critical region ($\varepsilon \leq 0.1$) can be expressed as
\begin{equation}
M(\mu_0H_\textrm{int},\varepsilon) = \varepsilon^\beta f_\pm(\mu_0H_\textrm{int}/\varepsilon^{\beta+\gamma}),
\end{equation}
where $f_-$ for $T<T_\textrm{c}$ and $f_+$ for $T>T_\textrm{c}$, respectively, are regular functions. Eq.(4) can be further written in terms of scaled magnetization $m\equiv\varepsilon^{-\beta}M(\mu_0H_\textrm{int},\varepsilon)$ and scaled field $\mu_0h\equiv\varepsilon^{-(\beta+\gamma)}\mu_0H_\textrm{int}$ as
\begin{equation}
m = f_\pm(\mu_0h).
\end{equation}
This suggests that for true scaling relations and the right choice of $\beta$, $\gamma$, $\delta$ values, the scaled $m$ and $\mu_0h$ will fall on universal curves above $T_\textrm{c}$ and below $T_\textrm{c}$, respectively.

\section{RESULTS AND DISCUSSION}

\begin{figure}
\centerline{\includegraphics[scale=1]{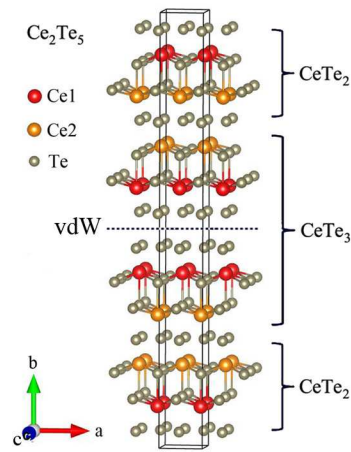}}
\caption{(Color online). Crystal structure of the van der Waals (vdW) magnet Ce$_2$Te$_5$ with alternating stacking of CeTe$_2$ and CeTe$_3$ along the $b$-axis.}
\label{XRD}
\end{figure}

\begin{figure}
\centerline{\includegraphics[scale=1]{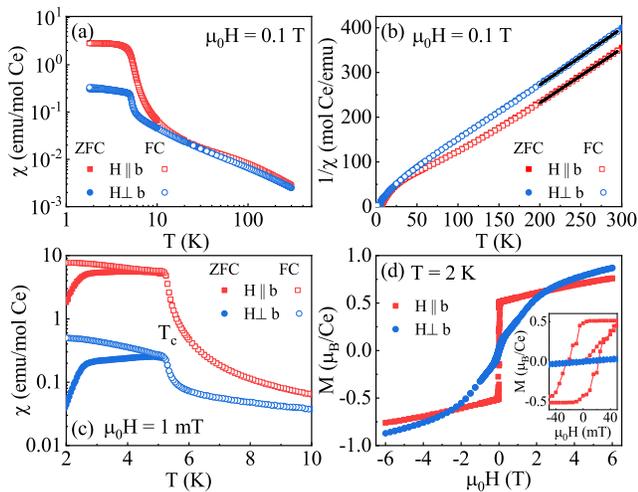}}
\caption{(Color online). Temperature dependence of (a) magnetic susceptibility $\chi$(T), defined as $M/\mu_0H$, and (b) inverse magnetic susceptibility $1/\chi$(T) of Ce$_2$Te$_5$ measured in magnetic field of $\mu_0H$ = 0.1 T applied parallel and perpendicular to the $b$-axis in zero-field-cooled (ZFC) and field-cooled (FC) modes. The solid lines are linear fits to the data. (c) The low temperature $\chi$(T) measured in a low magnetic field of $\mu_0H$ = 1 mT. (d) Field dependence of magnetization $M$($\mu_0H$) of Ce$_2$Te$_5$ measured at $T$ = 2 K.}
\label{MTH}
\end{figure}

Figure 2(a) shows the temperature dependence of magnetic susceptibility $\chi$(T) measured in $\mu_0H$ = 0.1 T applied parallel and perpendicular to the $b$-axis. A rapid upturn in $\chi$(T) at low temperature is observed for both field directions, indicating a FM transition. In this field, the zero-field-cooled (ZFC) and field-cooled (FC) data overlap well for each orientation. The temperature dependence of inverse susceptibility $1/\chi$(T) is plotted in Fig. 2(b). A linear fit from 200 to 300 K to a Curie-Weiss form, $\chi = C/(T-\theta)$, where $C$ is the Curie constant and $\theta$ is the paramagnetic Curie-Weiss temperature, gives $\theta$ = 9.6 K for $H \parallel b$ and -18.0 K for $H \perp b$, respectively. The positive value of $\theta$ for $H \parallel b$ is consistent with a dominant FM interaction, while the negative $\theta$ for $H \perp b$ suggests a dominant AFM interaction. The derived effective moment $\mu_{\textrm{eff}}$ = 2.56 $\mu_\textrm{B}/$Ce for $H \parallel b$ and 2.52 $\mu_\textrm{B}/$Ce for $H \perp b$, respectively, are very close to Hund's value for Ce$^{3+}$ of 2.54 $\mu_\textrm{B}$. It should be noted that the high-temperature anisotropy and deviation from Curie-Weiss behavior with decreasing temperature may be attributed to a crystalline electric field (CEF) effect as explained below. Figure 2(c) shows the low-temperature $\chi$(T) measured in a small field of $\mu_0H$ = 1 mT. In the ordered state, $\chi$(T) for $H \parallel b$ is 20 times larger than that of $H \perp b$, indicating a large magnetic anisotropy. The bifurcation of ZFC and FC curves below $T_\textrm{c} \approx 5.2$ K is likely due to a FM domain effect. Figure 2(d) displays the isothermal magnetization measured at 2 K. The magnetization $M(H \parallel b)$ rapidly saturates to $\sim$ 0.5 $\mu_\textrm{B}/$Ce at 40 mT, whereas $M(H \perp b)$ gradually increases. It is interesting that $M(H \perp b)$ increases up to a higher value than $M(H \parallel b)$ above 2.5 T [Fig. 2(d)], in line with the previous results \cite{Chen}, indicating that the magnetic order is more complex than simple ferromagnetism in Ce$_2$Te$_5$. A similar feature was also observed in bulk CrI$_3$ \cite{YL001}. As shown in the inset of Fig. 2(d), a clear hysteresis loop with coercive field $\mu_0H_\textrm{c} \approx 20$ mT is observed for $H \parallel b$, indicating soft ferromagnetism with an easy $b$ axis.

\begin{figure}
\centerline{\includegraphics[scale=1]{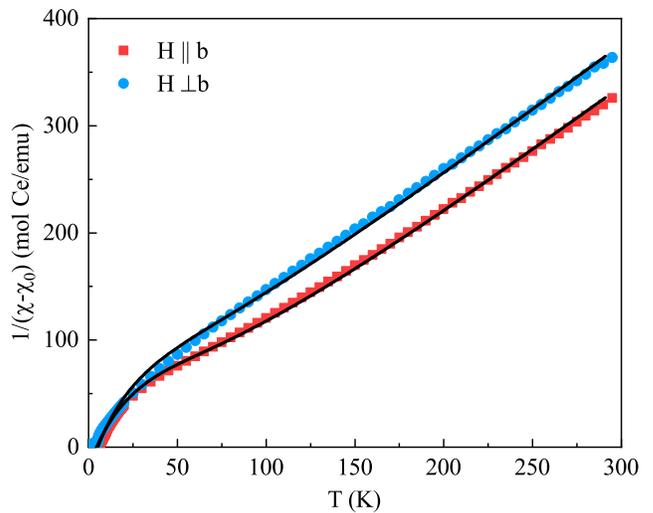}}
\caption{(Color online). Temperature dependence of $1/(\chi-\chi_0)$(T) of Ce$_2$Te$_5$ with a crystalline electric field (CEF) model fit (solid lines) as explained in the main text.}
\label{MTH}
\end{figure}

As shown in Fig. 3, the anisotropic magnetic susceptibilities can be modeled with a CEF Hamiltonian for Ce$^{3+}$ of total angular momentum $J = 5/2$. In Ce$_2$Te$_5$, due to $a$ and $c$ being accidentally degenerate in the orthorhombic $Cmcm$ crystal structure, the local Ce point group has $C_{4v}$ symmetry with the four-fold rotational axis along the crystallographic $b$-axis. The resultant CEF Hamlitonian contains three CEF parameters $B_n^m$ with corresponding Steven's operators $\hat{O}_m^n$ \cite{StevensOps} as
\begin{equation}
	H_{\textrm{CEF}} = B_2^0 \hat{O}_2^0 + B_4^0 \hat{O}_4^0 + B_4^4 \hat{O}_4^4,
\end{equation}
which produces three Kramers doublets. In the $C_{4v}$ point group and $J, m_j$ basis, these doublets are labeled $\Gamma_6$ with $m_j = \pm 1/2$ components or $\Gamma_7^1$/$\Gamma_7^2$ with mixed $m_j = \pm 5/2$ and $m_j = \pm 3/2$ components.
Magnetic susceptibility of the CEF Hamiltonian was calculated using Mantid Plot \cite{MP} with an additional temperature independent $\chi_0$ term and effective mean-field exchange interactions $\Theta_{\perp}$ and $\Theta_{\parallel}$. An effective susceptibility was calculated as $\chi_{\textrm{eff}} = \chi_{\textrm{CEF}}^{\textrm{calc}} / (1 - \Theta \chi_{\textrm{CEF}}^{\textrm{calc}})$ and compared to the observed $\chi_{\textrm{obs}} =  \chi - \chi_0$, where $\chi_0$ is a temperature-independent contribution. The overall fit was determined by minimizing $X^2 = (\chi_{\textrm{calc}} - \chi_{\textrm{obs}})^2/\chi_{\textrm{calc}}$ with a final $X^2 = 15.8$ and extracted parameters $B_2^0 = -0.39003$ meV, $B_4^0 = 0.06595$ meV, $B_4^4 = -0.21987$ meV, $\chi_0 = -0.000195$ emu mol$^{-1}$, $\Theta_{\perp} =24.74$ K, and $\Theta_{\parallel} =18.81$ K. The ground state doublet is $\Gamma_7^1 = 0.396|\pm 5/2 \rangle + 0.918|\mp 3/2 \rangle $, the first excited state doublet $\Gamma_7^2 = 0.918|\pm 5/2 \rangle - 0.396|\mp 3/2 \rangle$ is at 16.23 meV, and the second excited state doublet $\Gamma_6 = |\pm 1/2\rangle$ is at 24.67 meV. The ground state doublet has projected $g$ factors $g_{\perp} = 1.393$ and $g_{\parallel} = 1.497$, respectively. Taking $J_{\textrm{eff}} = 1/2$ for the ground state doublet, the expected saturated moment $gJ\mu_\textrm{B}$ is 0.697 $\mu_\textrm{B}$/Ce for $H \perp b$ and 0.749 $\mu_\textrm{B}$/Ce for $H \parallel b$, in reasonable agreement with the magnetization at 2 K and 6 T $M_{H \perp b}$ = 0.87 $\mu_\textrm{B}$/Ce and  $M_{H \parallel b}$ = 0.76 $\mu_\textrm{B}$/Ce.

\begin{figure}
\centerline{\includegraphics[scale=1]{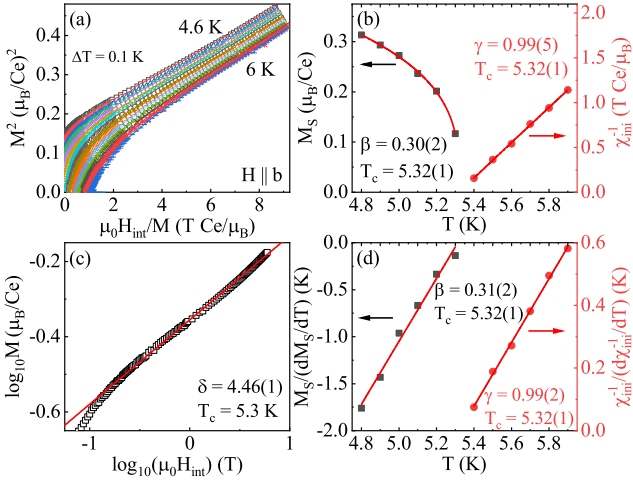}}
\caption{(Color online). (a) Arrott plots of $M^2$ vs $\mu_0H_\textrm{int}/M$ around $T_c$ for Ce$_2$Te$_5$ with $H \parallel b$. (b) Temperature dependence of the spontaneous magnetization $M_\textrm{s}$ (left) and the inverse initial susceptibility $\chi_{\textrm{ini}}^{-1}$ (right) with fitted curves for Ce$_2$Te$_5$ explained in the main text. (c) Isotherm in log$_{10}$M-log$_{10}$($\mu_0$$H_\textrm{int}$) measured at $T_\textrm{c}$ = 5.3 K, along with a linear fit to the data. (d) Kouvel-Fisher plots of $M_\textrm{s}(dM_\textrm{s}/dT)^{-1}$ (left) and $\chi_{\textrm{ini}}^{-1}(d\chi_{\textrm{ini}}^{-1}/dT)^{-1}$ (right), along with linear fits to the data.}
\label{Arrot}
\end{figure}

In order to understand the nature of the FM transition in Ce$_2$Te$_5$, one approach is to study in detail the critical exponents around $T_\textrm{c}$. Magnetization isotherms along the easy $b$-axis were measured from 4.6 to 6 K at intervals of 0.1 K. An Arrott plot of $M^2$ vs $\mu_0H_\textrm{int}/M$ at various temperatures is displayed in Fig. 4(a). In the mean field description of the magnetization near $T_\textrm{c}$, curves in the Arrott plot should be a series of parallel straight lines with the one passing through the origin indicating the $T_\textrm{c}$ \cite{Arrott1,Banerjee,Arrott2}. It is clear that mean field critical exponents do not work for Ce$_2$Te$_5$, as illustrated by a set of curved lines shown in Fig. 4(a). According to the Arrott-Noaks equation of state $(\mu_0H_\textrm{int}/M)^{1/\gamma} = a\varepsilon+bM^{1/\beta}$ \cite{Arrott2}, $a$ and $b$ are constants, a modified Arrott plot should be used to obtain the critical exponents. $\beta$ and $\gamma$ can be obtained self consistently. After selecting $\beta$ and $\gamma$, the linear extrapolation from the high field region to the intercepts with the axes $M^{1/\beta}$ and $(\mu_0H_\textrm{int}/M)^{1/\gamma}$ yields the values of $M_\textrm{s}$(T) and $\chi_{\textrm{ini}}^{-1}$(T). A new set of $\beta$ and $\gamma$ can be obtained by fitting data following Eqs. (1) and (2), which can be used to reconstruct a new modified Arrott plot. This procedure is then repeated until the values of $\beta$ and $\gamma$ are stable \cite{Pramanik}.

Figure 4(b) presents the final $M_\textrm{s}$(T) and $\chi_{\textrm{ini}}^{-1}$(T) with the fitted curves. Critical exponents $\beta = 0.30(2)$, $\gamma = 0.99(5)$, and $T_\textrm{c} = 5.32(1)$ K are obtained. Figure 4(c) exhibits the field dependence of magnetization of Ce$_2$Te$_5$ at $T_\textrm{c} = 5.3$ K in a log$_{10}$M-log$_{10}$($\mu_0$$H_\textrm{int}$) plot, yielding $\delta$ = 4.46(1) from Eq. (3). In comparison to the theoretical prediction based on the Widom relation \cite{Kadanoff},
\begin{equation}
\delta = 1+\frac{\gamma}{\beta}
\end{equation}
the derived $\delta$ = 4.3(1) is close to that obtained in Fig. 4(c). In addition, critical exponents can also be determined according to the Kouvel-Fisher (KF) method \cite{Kouvel}:
\begin{equation}
\frac{M_\textrm{s}(T)}{dM_\textrm{s}(T)/dT} = \frac{T-T_\textrm{c}}{\beta}
\end{equation}
\begin{equation}
\frac{\chi_{\textrm{ini}}^{-1}(T)}{d\chi_{\textrm{ini}}^{-1}(T)/dT} = \frac{T-T_\textrm{c}}{\gamma}
\end{equation}
$M_\textrm{s}(T)/[dM_\textrm{s}(T)/dT]$ and $\chi_{\textrm{ini}}^{-1}(T)/[d\chi_{\textrm{ini}}^{-1}(T)/dT]$ are linear functions of temperature with slopes of $1/\beta$ and $1/\gamma$, respectively. As shown in Fig. 4(d), the linear fits give $\beta = 0.31(2)$, $\gamma = 0.99(2)$, and $T_\textrm{c} = 5.32(1)$ K, which are consistent with those generated by the modified Arrott plot.

\begin{figure}
\centerline{\includegraphics[scale=1]{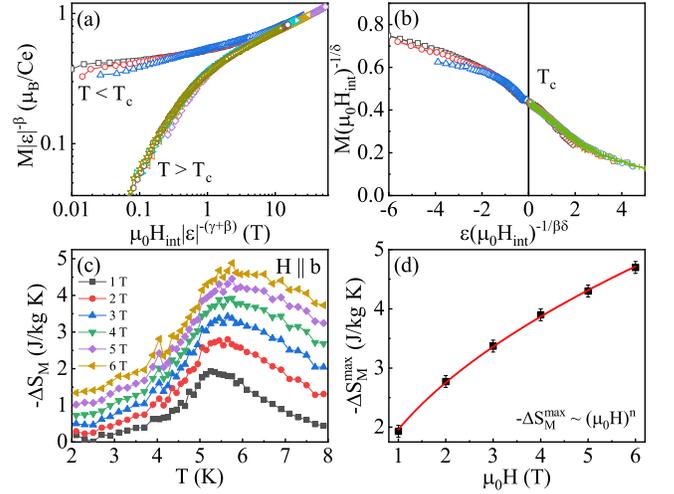}}
\caption{(Color online). (a) The scaling plots of normalized magnetization $m=M|\varepsilon|^{-\beta}$ as a function of normalized field $\mu_0h=\mu_0H_\textrm{int}|\varepsilon|^{-(\gamma+\beta)}$ below and above $T_c$ for Ce$_2$Te$_5$. (b) Scaled magnetization $M(\mu_0H_\textrm{int})^{-1/\delta}$ vs scaled field $\varepsilon (\mu_0H_\textrm{int})^{-1/\beta\delta}$. (c) The magnetic entropy change $-\Delta S_M$ obtained from magnetization at various magnetic fields along the $b$-axis. (d) Field dependence of the maximum magnetic entropy change $-\Delta S_M^{max}$ with a power law fit.}
\label{3D}
\end{figure}

Following Eq. (5), the scaled $m$ vs scaled $\mu_0h$ is plotted in Fig. 5(a). All the data reasonably well collapse into two separate branches: one below $T_\textrm{c}$ and another above $T_\textrm{c}$. The scaling equation of state also takes another form:
\begin{equation}
\frac{\mu_0H_\textrm{int}}{M^\delta} = k[\frac{\varepsilon}{(\mu_0H_\textrm{int})^{1/\beta}}]
\end{equation}
where $k(x)$ is the scaling function. Figure 5(b) shows $M(\mu_0H_\textrm{int})^{-1/\delta}$ vs $\varepsilon (\mu_0H_\textrm{int})^{-1/(\beta\delta)}$ for Ce$_2$Te$_5$, where the experimental data collapse reasonably onto a single curve, and $T_\textrm{c}$ locates at the zero point of the horizontal axis. The well-scaled curves confirm reliability of the obtained critical exponents. Figure 5(c) presents the derived magnetic entropy change $-\Delta S_\textrm{M} = \int_0^{\mu_0H} \left[\partial M(T,\mu_0H)/\partial T\right]_{\mu_0H}d(\mu_0H)$ \cite{Amaral}, which shows a broad peak centered near $T_\textrm{c}$. The peak value monotonically increases with increasing field, reaches 4.7(1) J kg$^{-1}$ K$^{-1}$ in 6 T. The field dependence of $-\Delta S_\textrm{M}^{\textrm{max}}$ follows a power law $-\Delta S_\textrm{M}^{\textrm{max}} \propto (\mu_0H)^n$ with $n = 1+(\beta-1)/(\beta+\gamma)$ \cite{VFranco,YLMC1,YLMC2,YLMC3,YLMC4}. Fitting of $-\Delta S_\textrm{M}^{\textrm{max}}$ gives $n = 0.49(1)$, which is close to the calculated value of 0.46(1), further verifies the reliability of the obtained critical exponents.

\begin{table*}
\caption{\label{tab1}Comparison of critical exponents of Ce$_2$Te$_5$ with different theoretical models. MAP, KFP, and CI represent the modified Arrott plot, the Kouvel-Fisher plot, and the critical isotherm, respectively.}
\begin{ruledtabular}
\begin{tabular}{lllllll}
   & Reference & Technique & $T_\textrm{c}$ & $\beta$ & $\gamma$ & $\delta$ \\
  \hline
  Ce$_2$Te$_5$ & This work & MAP & 5.32(1) & 0.30(2) & 0.99(5) & 4.30(5)\\
  &  & KFP & 5.32(1) & 0.31(2) & 0.99(2) & 4.2(1) \\
  &  & CI  &  5.3  &   &   & 4.46(1) \\
  2D Ising & \cite{Widom} & Theory & & 0.125 & 1.75 & 15.0 \\
  Mean field & \cite{Kouvel} & Theory & & 0.5 & 1.0 & 3.0 \\
  3D Heisenberg & \cite{Kouvel} & Theory & & 0.365 & 1.386 & 4.8 \\
  3D Ising & \cite{Huang3} & Theory & & 0.325 & 1.24 & 4.82 \\
  3D XY & \cite{Phan} & Theory & & 0.345 & 1.316 & 4.81 \\
  Tricritical mean field & \cite{Fisher1972} & Theory & & 0.25 & 1.0 & 5.0
\end{tabular}
\end{ruledtabular}
\end{table*}

Taroni \emph{et al.} pointed out that the value of $\beta$ for a 2D magnet should be within a window $0.1 \leq \beta \leq 0.25$ \cite{Taroni}. The value of $\sim$ 0.31 obtained here indicates a clear 3D behavior in Ce$_2$Te$_5$. As we can see, the critical exponent $\beta$ of Ce$_2$Te$_5$ is close to the theoretical value ($\beta = 0.325$) of the 3D Ising model (Table I), consistent with the large anisotropy in magnetization below $T_\textrm{c}$ [Fig. 2(d)]. However, the value of $\gamma \sim 1.0$ of Ce$_2$Te$_5$ deviates from $\gamma = 1.24$ of the 3D Ising model, which might be arising from the long-range Ruderman-Kittel-Kasuya-Yosida (RKKY) interactions.

\begin{figure}
\centerline{\includegraphics[scale=1]{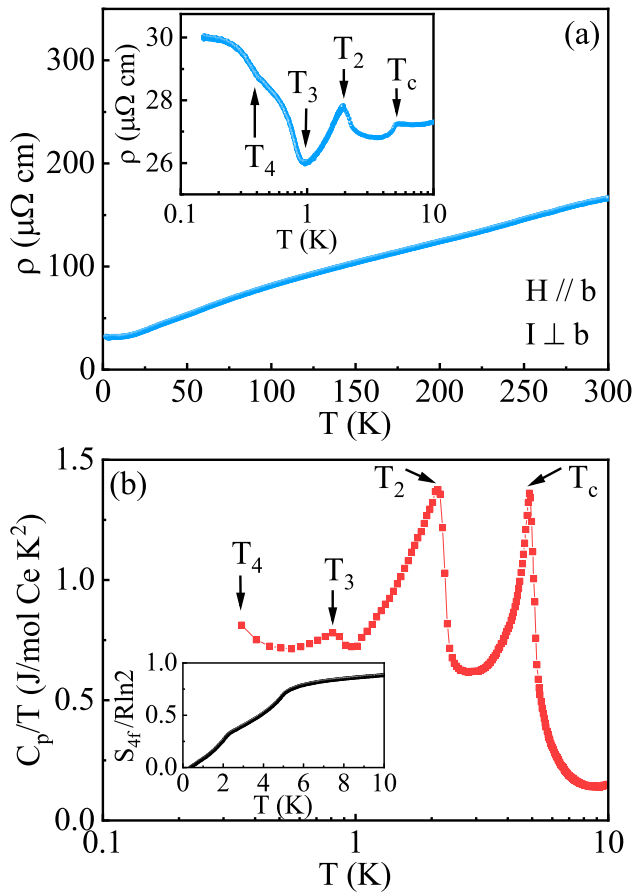}}
\caption{(Color online). Temperature dependence of (a) in-plane electrical resistivity $\rho$(T) and (b) specific heat $C_\textrm{p}/T$ of Ce$_2$Te$_5$ in zero field. Inset in (b) shows the electronic entropy $S_{4f}$(T)/Rln2 of Ce$_2$Te$_5$.}
\label{NS}
\end{figure}

Having delineated salient features of the FM state below $T_\textrm{c}$, we now turn to an investigation of electrical transport properties. Figure 6(a) shows the temperature dependence of $ac$-plane resistivity $\rho$(T) of Ce$_2$Te$_5$, a typical metallic behavior with slope change around 100 K. This might be related to the hybridization between local moments and conduction electrons, and the presence of a CEF doublet near 190 K. The low-temperature $\rho$(T) of Ce$_2$Te$_5$ displays four consecutive transitions at 5.2, 2.1, 0.9, and 0.4 K [inset in Fig. 6(a)]. The first transition temperature corresponds well to the FM transition ($T_\textrm{c}$) in $\chi$(T) [Fig. 2(a)]. Below $T_\textrm{c}$ and $T_2$, $\rho$(T) decreases more rapidly due to a decrease of spin disorder scattering; however, $\rho$(T) shows a rapid upturn below $T_3$ and a weak kink at $T_4$ that suggest the opening of a small gap possibly due to a spin-density-wave. The specific heat divided by temperature, $C_\textrm{p}/T$ [Fig. 6(b)], shows peaks at 5.2 K, 2.1 K, 0.9 K, and an upturn toward 0.4 K that are consistent with the four transitions observed in $\rho$(T). The electronic entropy $S_{4f}$ [inset in Fig. 6(b)] is determined by subtracting a $\beta T^3$ phonon contribution from $C_\textrm{p}/T$ (obtained from a linear fit to $C_\textrm{p}/T = \gamma + \beta T^2$ from 9 to 10 K), and integration of the resulting $4f$ contribution to the specific heat $S_{4f} = \int{(C_{4f}/T)}dT$. The value of $S_{4f}$ approaches $R$ln2 at 10 K, indicating a crystal field doublet ground state in Ce$_2$Te$_5$. About 75\% of $R$ln2 of entropy is released below $T_\textrm{c}$ and about 30\% of $R$ln2 is released below $T_2$. One plausible scenario is that one Ce sublattice orders ferromagnetically below $T_\textrm{c}$, while the other Ce sublattice orders antiferromagnetically below $T_2$. This scenario is consistent with the roughly equal amount of entropy released below $T_\textrm{c}$ and $T_2$ [Fig. 6(b) inset], the soft ferromagnetic response of $M$(H) at $T$ = 2 K below $T_2$, the field dependence of the ordering temperatures discussed below, and the competition of magnetic interactions in Ce$_2$Te$_5$. Alternatively, magnetic coupling between the two magnetic Ce sublattices may give rise to interesting magnetically ordered states, such as ferrimagnetism below $T_c$, coexistent FM and AFM states at 2 K (see Fig. 2d), or incommensurate ordered states. Further neutron scattering $\mu$SR measurements will be useful to determine the nature of the magnetically ordered states and their underlying magnetic interactions in Ce$_2$Te$_5$.

With increasing magnetic fields applied along the $b$ axis, as shown in Fig. 7(a), the first transition $T_\textrm{c}$ broadens and shifts to higher temperature, confirming it is a FM transition. $T_2$ and $T_3$, however, move to lower temperature and are finally suppressed in high fields, indicating an AFM character. Figure 7(b) shows the field dependence of longitudinal magnetoresistance (MR) measured at 0.5, 1.5 and 4 K. The MR at 4 K shows a relatively large positive value of 75\% in $\mu_0H$ = 9 T. In addition, an inflection was observed at low temperatures. As depicted in the inset of Fig. 7(b), a peak in $d$MR/$d(\mu_0H)$ indicates that a critical field of $\sim$ 4.1 T at 0.5 K and 3.5 T at 1.5 K, respectively, which are associated with the $T_2$ and $T_3$ transitions. Based on these transport measurements, the magnetic phase ($T-\mu_0H$) diagram of Ce$_2$Te$_5$ is summarized in Fig. 7(c). The contribution of the interlayer magnetic interactions between CeTe$_2$ and CeTe$_3$ likely generates complex magnetic interactions in Ce$_2$Te$_5$. The nature of $T_4$ at the lowest temperature needs further investigation, and neutron scattering measurements are required to determine the CEF levels and the magnetic structure of Ce$_2$Te$_5$.

\begin{figure}
\centerline{\includegraphics[scale=1]{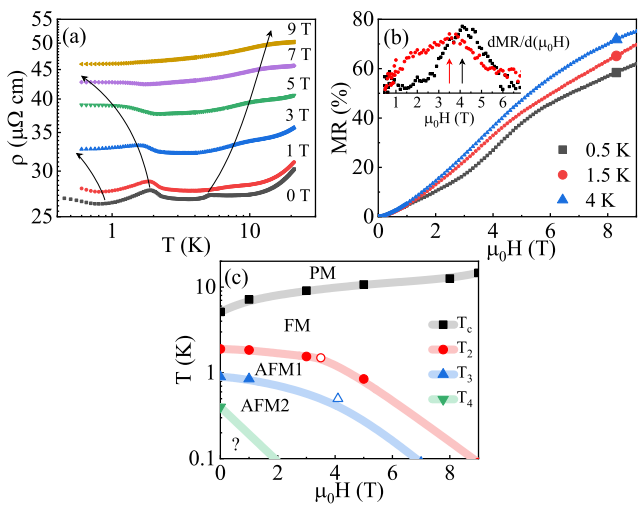}}
\caption{(Color online). (a) Magnetic-field variations of in-plane electrical resistivity $\rho$(T) for $I \perp b$ and $H \parallel b$ for a single crystal of Ce$_2$Te$_5$. The lines are guides to the eye. (b) The longitudinal magnetoresistance (MR) of Ce$_2$Te$_5$ at indicated temperatures. Inset shows $d$MR/$d(\mu_0H)$ curves at 0.5 and 1.5 K. (c) The magnetic phase diagram constructed from transport measurements (solid symbols from $\rho$; open symbols from MR). PM, FM, and AFM represent paramagnetic, ferromagnetic, and antiferromagnetic phases, respectively. The lines are guides to the eye.}
\label{NS}
\end{figure}

\begin{figure}
\centerline{\includegraphics[scale=1]{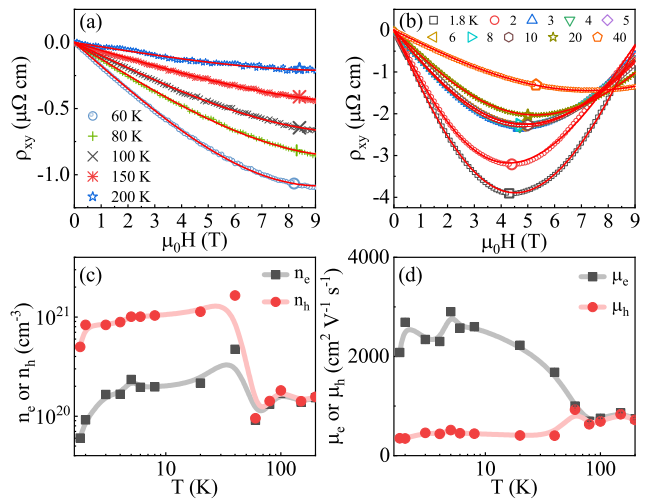}}
\caption{(Color online). The field dependence of Hall resistivity $\rho_{xy}$($\mu_0H$) at various (a) high and (b) low temperatures with $I \perp b$ and $H \parallel b$ for a single crystal of Ce$_2$Te$_5$. The lines in (a) and (b) are fits of the two-band model to the data as explained in the main text. The derived (c) electron concentration $n_e$ and hole concentration $n_h$, (d) electron mobility $\mu_e$ and hole mobility $\mu_h$.}
\label{NS}
\end{figure}

To shed light on the transport carriers in Ce$_2$Te$_5$, we further measured the field dependence of Hall resistivity $\rho_{\textrm{xy}}$($\mu_0H$) of Ce$_2$Te$_5$ with current flowing in the $ac$-plane and the field applied along the $b$-axis at various temperatures, as shown in Figs. 8(a) and 8(b). The Hall coefficient $\rho_{\textrm{xy}}$/($\mu_0H$) is negative at high temperatures, indicating dominant electron-like carriers. With decreasing temperature, $\rho_{\textrm{xy}}$ exhibits nonlinear behavior below 60 K, and the shape of $\rho_{\textrm{xy}}$ changes significantly below 40 K, becoming parabolic at low temperatures. These observations suggest that Ce$_2$Te$_5$ is a multi-band system, as has been observed in a related compound GdTe$_3$ \cite{Lei}. Assuming a two-band model, the Hall resistivity is expressed as,
\begin{equation}
\rho_{\textrm{xy}} = \frac{\mu_0H}{e}\frac{(n_h\mu_h^2-n_e\mu_e^2)+(n_h-n_e)(\mu_e\mu_h\mu_0H)^2}{(n_h\mu_h+n_e\mu_e)^2+(n_h-n_e)^2(\mu_e\mu_h\mu_0H)^2}
\end{equation}
where $e$ is the elementary charge; $n_e$ and $n_h$ are the electron and hole carrier concentrations, respectively; $\mu_e$ and $\mu_h$ are the electron and hole carrier mobilities, respectively. The derived fitting parameters $n_e$, $n_h$, $\mu_e$ and $\mu_h$ are plotted in Figs. 8(c) and 8(d), respectively, where the electron and hole carriers are almost compensated at high temperatures and become uncompensated below 60 K. The carrier concentration is of the order of 10$^{20}$ - 10$^{21}$ cm$^{-3}$ [Fig. 7(c)]. The hole mobility $\mu_h$ is $\sim$ 350 - 930 cm$^2$/V$\cdot$s and is weakly temperature-dependent; however, $\mu_e$ increases abruptly below 60 K and features a relatively large value of $\sim$ 2900 cm$^2$/V$\cdot$s near $T_\textrm{c}$. This high electron mobility in Ce$_2$Te$_5$, a vdW layered and magnetically ordered material, is larger than that of some other rare-earth materials such as YbMn(Bi,Sb)$_2$ and EuMnBi$_2$ \cite{Wang,Wang1,May}, and provides further opportunities for future device applications.

\section{CONCLUSIONS}

In summary, we studied the magnetic, transport, and thermodynamic properties of the $f$-electron vdW magnet Ce$_2$Te$_5$ and summarized its magnetic phase diagram. Four magnetic transitions were observed at $T_\textrm{c}$ = 5.2, $T_2$ = 2.1, $T_3$ = 0.9, and $T_4$ = 0.4 K. Critical exponents in the vicinity of the FM transition are determined to be $\beta = 0.31(2)$, $\gamma = 0.99(2)$, $\delta = 4.46(1)$, $T_\textrm{c} = 5.32(1)$ K, indicating that Ce$_2$Te$_5$ shows a three-dimensional magnetic critical behavior. A crystal electric field model of the magnetic susceptibility suggests that the ground state $\Gamma_7^1$ doublet has a dominant $|3/2 \rangle $ character with excited $\Gamma_7^2$ and $\Gamma_6$ states at $\sim$ 16 and 25 meV, respectively. Further neutron scattering or X-ray absorption spectroscopy measurements will be needed to confirm this CEF scheme. The Hall resistivity analysis indicates that Ce$_2$Te$_5$ is a multi-band system with a relatively high electron mobility around $T_\textrm{c}$. Furthermore, with rapid developments in the field of 2D materials, we expect our experimental work to stimulate broad interest for exploring its magnetic and transport properties in the 2D limit.

\section*{Acknowledgements}

Work at Los Alamos National Laboratory was performed under the auspices of the U.S. Department of Energy, Office of Basic Energy Sciences, Division of Materials Science and Engineering under project ``Quantum Fluctuations in Narrow-Band Systems". Y.L., M.M.B., and A.W. acknowledges the Director's Postdoctoral Fellowship through the Laboratory Directed Research and Development program.

\end{document}